\documentclass[11pt]{article}
\usepackage{amssymb,amsmath,amsfonts,amsthm,mathtools,color}

\usepackage{mathrsfs}
\usepackage{bm} 

\usepackage{comment} 
\usepackage[title,titletoc,header]{appendix}
\usepackage{graphicx,subfig}

\usepackage{paralist}

\usepackage{tikz}
\usetikzlibrary{arrows,positioning,shapes.geometric}

\usepackage[linesnumbered, ruled,vlined]{algorithm2e}
\SetKwInput{KwInput}{Input}                

\graphicspath{ {./figures/} }

\usepackage{indentfirst}
\usepackage{multicol}
\usepackage{booktabs}
\usepackage{url}
\usepackage[outdir=./]{epstopdf} 

\usepackage[shortlabels]{enumitem}

\usepackage{hyperref}
\hypersetup{
    colorlinks=true, 
    linktoc=all,     
    linkcolor=blue,  
}
\numberwithin{equation}{section}
\usepackage{multirow}

\theoremstyle{definition}

\theoremstyle{remark}
\newtheorem{Remark}{Remark}[section]

\def\to{\rightarrow}

 \def\t{\times}

\def\cA{\mathcal{A}}

\def\cH{\mathcal{H}}

\def\cP{\mathcal{P}}

\def\cX{\mathcal{X}}

\def\sE{{\mathbb{E}}}

\def\sN{{\mathbb{N}}}

\def\sR{{\mathbb R}}

\DeclareMathOperator*{\argmin}{arg\,min}

\newcommand{\lc}
{\mathrel{\raise2pt\hbox{${\mathop<\limits_{\raise1pt\hbox
{\mbox{$\sim$}}}}$}}}

\newcommand{\gc}
{\mathrel{\raise2pt\hbox{${\mathop>\limits_{\raise1pt\hbox{\mbox{$\sim$}}}}$}}}

\newcommand{\ec}
{\mathrel{\raise2pt\hbox{${\mathop=\limits_{\raise1pt\hbox{\mbox{$\sim$}}}}$}}}

\def\bb{\begin{equation}} \def\ee{\end{equation}}
\def\bbn{\begin{equation*}} \def\een{\end{equation*}}

\def\beqn{\begin{eqnarray}}  \def\eqn{\end{eqnarray}}

\def\beqnx{\begin{eqnarray*}} \def\eqnx{\end{eqnarray*}}

\def\bn{\begin{enumerate}} \def\en{\end{enumerate}}

\def\bd{\begin{description}} \def\ed{\end{description}}



\title{Insurance pricing
on price comparison websites via 
Reinforcement Learning}

\author{
\and Tanut Treetanthiploet\thanks{Quantum Technology Foundation,  \texttt{ttreetanthiploet@gmail.com}}
\and 
Yufei Zhang\thanks{Department of Statistics, London School of Economics and Political Science,  \texttt{y.zhang389@lse.ac.uk}}
\and 
 Lukasz Szpruch\thanks{School of Mathematics, University of Edinburgh and Alan Turing Institute,  \texttt{L.Szpruch@ed.ac.uk}}
 \and 
 Isaac Bowers-Barnard\thanks{Accenture (UK) Limited,  \texttt{isaac.bowers-barnard@mudano.com, henrietta.ridley@accenture.com  }}
 \and
 Henrietta Ridley\footnotemark[4]
 \and
 James Hickey\thanks{esure Services Limited,  \texttt{james.hickey@esure.com, Chris.Pearce@esure.com}}
 \and
 Chris Pearce\footnotemark[5]
}

\begin{document}

\maketitle

\begin{abstract}
  The emergence of price comparison websites (PCWs) has presented insurers with unique challenges in formulating effective pricing strategies. Operating on PCWs requires insurers to strike a delicate balance between competitive premiums and profitability, amidst obstacles such as low historical conversion rates, limited visibility of competitors' actions, and   a dynamic market environment. In addition to this, the capital intensive nature of the business means pricing below the risk levels of customers can result in solvency issues for the insurer.
To address these challenges, this paper introduces reinforcement learning (RL) framework that learns the optimal pricing policy by integrating model-based and model-free methods. 
The model-based component is used to train agents in an offline setting, avoiding cold-start issues, while model-free algorithms are then employed in a contextual bandit (CB) manner to dynamically update the pricing policy to maximise the expected revenue. This  facilitates quick adaptation to evolving market dynamics and 
 enhances algorithm efficiency and decision  interpretability.
The paper also highlights the importance of evaluating pricing policies using an offline dataset in a consistent fashion and demonstrates the superiority of the proposed methodology over existing off-the-shelf RL/CB approaches.
We validate our methodology using synthetic data, generated to reflect private commercially available data within real-world insurers, and compare against $6$ other benchmark approaches. Our hybrid agent outperforms these benchmarks in terms of sample efficiency and cumulative reward with the exception of an agent that has access to perfect market information which would not be available in a real-world set-up.

\end{abstract}


\section{Introduction}

The rise of price comparison websites (PCWs) has transformed the general insurance industry in the UK, granting consumers extensive access to diverse insurance options, particularly, for home and motor insurances. These platforms consolidate premiums, coverage details, and policy terms from multiple insurers, enabling users to make informed decisions. This has intensified competition among insurers, compelling them to offer competitive pricing and attractive policy features. Online insurance pricing has become an  indispensable component of insurers' business strategies, shaping their market presence and overall success. Current industry standards are to leverage supervised learning models, trained offline on historic data, to feed into an online optimiser. This can lead to a lot of maintenance as the system requires many models to be maintained and keeping pace with the dynamic market is difficult as information relating to market prices typically becomes available weeks after a quote. At this point, the pricing behaviours of the market may have drifted significantly - for example, it is not uncommon for insurers to perform multiple targeted price changes within a single week.

\paragraph{Objectives and challenges in online insurance pricing.} 

Insurers operating on price comparison platforms face unique challenges in determining online insurance pricing.
These challenges stem from the delicate task of finding the right balance between offering competitive premiums and maintaining profitability,  while taking into account customer preferences \cite{verschuren2022,guelman2014},  market dynamics, and regulatory obligations.
Setting a high premium may increase revenue but risk\textcolor{violet}{s}  customer rejection, while offering a lower premium than competitors may boost conversion rates on quotes but negatively impact profitability and solvency.
The fact that insurers lack direct visibility into competitors' pricing strategies and the customers' price sensitivity at point of quote
further complicates the search for the optimal pricing rule. 

This paper studies   the online pricing problem faced by an individual insurer. This problem can be described as a sequential decision process as follows.
During a specific period, multiple customers arrive sequentially at the PCW. Each customer is characterised by a feature vector that includes information like age and claims information. The insurer determines the price to submit to the platform based on customer features, considering both the insurer's constraints and objectives as well as estimates of the customer's price preference \cite{verschuren2022}. Upon receiving quotes from all insurers,  the customer  decides whether to accept one of the offered prices. Market price coming from competitors are unknown at the point of quote. The latter information is available at a later date, post-policy start date for the customer, to avoid anti-competitive behaviours by insurers.  This information comes in the form of aggregate market prices for a quote, e.g., market quantiles, and estimates of this information are often used in place of the actual hidden values during price optimisation.

One of the core items used in this price determination, is the insurers estimate of the customer lifetime value (CLTV) given the details provided. Insurers typically estimate CLTV values at multiple time-horizons, with 1st-year CLTV (CLTV1) reflecting expected profit excluding profits that may arrive via renewals. The insurer's objective is to  devise an optimal pricing strategy that maximises the accumulated CLTV  from arriving customers within the specified time period subject to constraints such as a target conversion rate. Depending on the customer's decision, the insurer either receives a positive reward equal to the estimated CLTV for the quoted price or no reward at all. 

We propose an offline reinforcement learning (RL) algorithm  \cite{prudencio2023,fujimoto2019}  to learn  a pricing policy using  a static dataset comprising historical quoted prices and their corresponding outcomes. Unlike   traditional online RL approaches \cite{sutton2018reinforcement}, the proposed offline RL approach 
avoids directly interacting with the PCW, and 
learns  pricing strategies in a more cost-effective manner. Moreover, this offline approach enables a more flexible and controlled training/testing process compared to off-the-shelf methodologies.  Traditional online RL algorithms rely on posing quotes on   PCWs  and refining pricing rules based on customer responses. Given the competitive nature of the UK Insurance market, where conversion rates are low and a large number of insurers compete for every quote, this results in a sparse feedback signal and potentially catastrophic pricing policies at the early-stages of learning for fully online algorithms. During the initial training stage, these online algorithms often yield  inaccurate and unstable prices, posing significant financial and reputational risks for insurers. 
 
However, developing offline RL algorithms for insurance pricing encounters several challenges which we now describe.
(i) Sparse reward: 
The low conversion rate of insurance products results in a low signal-to-noise ratio in the data. In practice, a successful insurance product may have {less than} $2\%$ of quoted prices accepted by customers and generate positive rewards. This sparsity presents challenges for off-the-shelf RL algorithms, as they may struggle to train or require substantial amounts of data that may not be readily accessible.
(ii) Partial observability: Insurers have limited visibility into the actions of other insurers on the PCW. (iii) Non-stationarity: Real-world data shows that the performance of calibrated pricing rules deteriorates over time, often within weeks, due to market dynamics. As a result, it is crucial to develop algorithms that can adapt quickly to changing market scenarios.
(iv) Interpretability: ``Black-box" model-free RL algorithms may produce pricing rules that lack interpretability and fail to meet regulatory expectations. 

Although existing methodologies have been proposed in the literature to address individual aspects of these challenges for generic RL problems, they often require customisation to the specific problem setting for practical deployment. These customisations often take the form of novel reward design~\cite{eschmann2021}, architectures or training methodologies. Notably, there is a lack of published work  that  tailors these RL techniques specifically for the insurance pricing problem at hand. The main contribution in this areas focusses instead on pricing at renewals \cite{krashen2019}, formalising this setting as a constrained Markov decision problem with a coarse-coded state space. This setting is less competitive than the ``New Business" PCW driven framework we address, avoiding the described sparsity issues as renewals rates are much higher than conversion rates on PCWs.

 \paragraph{Our work.}
   
   This paper proposes a novel framework for training and evaluating RL algorithms in insurance pricing using historical data. The learning problem is first transformed into a contextual bandit (CB) problem \cite{collier2018}, assuming that customers arrive independently according to an unknown distribution, and insurer quotations and customer price responsiveness depend solely on customer provided features.  Within this set-up, we ignore additional constraints provided and instead focus on providing prices to maximise profitability. In this case, we measure profitability with CLTV1. The framework allows for constraints to be incorporated into the training paradigm. These can also be applied during live predictions where conversion rates are often aligned by insurers using percentage changes applied to the final prices produced during the roll-out of new pricing rules. This simplifies the training process and allows for focusing on maximising immediate rewards based on the current customer features with longer-term expectations of customer profitability built into the reward design. 
   
   We then introduce a novel hybrid algorithm to solve the contextual bandit problem. This algorithm combines model-based and model-free RL methods as follows: 
 \begin{itemize}
\item 
First, a conversion model is estimated for each customer feature and market quantile price(s), capturing the customer responsiveness to different quoted prices. This model is trained using historical data over a relatively long time horizon, exploiting the price-driven nature of the markets to capture stable pricing patterns.
The estimated conversion model plays a critical role in simulating customer reactions to prices quoted by RL algorithms, enabling effective algorithm training and evaluation. This alleviates issues surrounding limited exploration in historical batch-data that other approaches are adapted for \cite{fujimoto2019}.
\item 
Next, the reward is reformulated as the  \textit{expected} CLTV1 at the proposed price point, utilising the estimated conversion model. This transformation converts the sparse reward into a dense reward, enhancing  the sample efficiency of the  algorithm, and makes the system more interpretable \cite{devidze2021}. In this set-up the agent is trained to act so as to maximise the expected in-year profit as estimated by the insurers traditional views of risk, margin and conversion within the market. This is characterised by the reward which is determined by customer features used that are already live in existing pricing rules. This reduces the RL interpretability issue  to a  {model interpretability issue}, for which there are many paradigms~\cite{lundberg2017,shrikumar2017,saleem2022}, as all of the other components in the system design are currently used in existing pricing decisions and subject to interpretability constraints. Moreover,  this design decision eliminates potential issues arising from uncertainties in longer-term CLTV estimates and makes the agent more risk-averse as it does not leverage expected reward many years later which are less certain to arrive and depend on the market at the point in the future.

 \item 
Lastly, a model-free approach is used to sequentially update the pricing policy based on customer features sampled from the training dataset.
This eliminates the need to model the non-linear and non-stationary dependence  of the market quantile price(s) on customer features, distinguishing it from traditional model-based pricing algorithms (see \cite{blier2020machine}). 
It also allows for dynamically updating the pricing rule as new data arrives,
 making the pricing rule adapt quickly to changes of the market dynamics, such as competitors re-calibrating their methodologies.
  
\end{itemize}
To the best of our knowledge, this is the first offline RL framework that
 addresses insurance pricing problems on competitive price comparison platforms while tackling the practical challenges of sparse reward, partial observation, and non-stationary market environments. It accounts for the realities of data availability in this setting, through the use of an offline conversion model that leverages market data   available in the live setting, as well as increases transparency/interpretability through our specific reward design.  

We further extend the above methodology to systematically evaluate pricing policies generated by RL algorithms using an offline testing dataset prior to their actual deployment. By applying this methodology, we demonstrate the superiority of the proposed algorithm compared to several off-the-shelf fully model-based and fully model-free RL algorithms. 

The rest of this paper is organised as follows. Section \ref{sec:problem_formulation} formulates the insurance pricing as a reinforcement learning/contextual bandit problem. Section \ref{sec:RL_methodology} details the application of RL to this problem with numerical experiments on representative synthetic data, derived from real motor insurance quote data, provided in Section \ref{sec:experiments}. Finally, Section \ref{sec:conclusions} presents our conclusions and possible extensions and further work to build on our methodology.

\paragraph{Related works.}
RL has been applied to pricing problems
but the literature is limited and sparse. For instance, \cite{rana2015dynamic} and \cite{krashen2019}  employed RL techniques to adjust prices for interdependent perishable products and insurance products at renewals, respectively. \cite{raju2006learning} utilised RL to determine dynamic prices in an electronic retail market.  \cite{ban2021personalized, khraishi2022offline, luo2023distribution}  applied RL to price consumer credit. It is important to note that all of these studies focused on the single-agent pricing problem within a stationary environment. In contrast, our paper addresses a more challenging   pricing problem in a competitive multi-agent environment. Other related settings such as real-time bid-optimisation for online advertisements, often utilise a model-based approach \cite{cai2017}, rely on modifying existing strategies \cite{liu2022} or use transfer-learning of supervised methods to optimise the price \cite{han2021}. This last approach is very similar to the current standard for pricing engines within the market which is proving  insufficient given the dynamic nature of the environment. As alluded to earlier, 
adapting RL techniques to this setting poses unique challenges, such as the low signal-to-noise ratio, partial observation of competitors' actions, and the non-stationarity of the underlying environment.

 \section{Problem formulation}
 \label{sec:problem_formulation}

This section formulates  the new business insurance pricing problem as an offline learning problem.

Let
$\cX$ be the set of all customer features,
and 
  $\cA\subset (0,\infty)$ be the agent's action space.
  In practice, the set $\cA$ can either be   the agent's quoted price
  or  the ratio of the quoted price with respect to a reference price (i.e. benchmark premium). 
  In the latter instance, it is common for the action space to be discretised and to be restricted to a finite set of ratios in line with the insurer's discount/price increase appetite. 

Let $T\in \mathbb{N}$ be a given time horizon.
For each pricing policy  $\phi: \{1, ..., T\} \times \cX \to  \cA$,
the cumulative reward of the agent is  given by
\begin{equation}
\label{eq: RL objective}
    \sE \left[ \sum_{t=1}^T Y_t \,  r(X_t, \phi(t, X_t) ) 
    \right],
\end{equation}
where
$r(x,a)$  is the     lifetime value    for a  customer with   feature $x\in \cX$ when the price action $a\in \cA$ is offered,
 $X_t$ is  a random variable  representing the feature of the customer arrived at time $t$,
$\phi(t,X_t)$ is the price quoted by the agent at time $t$,
and $Y_t\in\{0,1\}$ is a random variable representing  the customer's decision,
i.e., $1$ indicates the agent's offer is accepted by the customer  and $0$ indicates the offer is  rejected.  
The agent  aims to learn  a  policy $\phi $ that maximises the cumulative reward \eqref{eq: RL objective}:
\begin{equation}
\label{eq: RL objective_optimal}
\phi^\star\in \argmin_{\phi:\{1, ..., T\} \times \cX}    \sE \left[ \sum_{t=1}^T Y_t \,  r(X_t, \phi(t, X_t) ) 
    \right].
\end{equation}

Note that in the above pricing problem, 
the customer's decision $Y_t$ depends on the customer information $X_t$,
the agent's action/quoted price $\phi(t,X_t)$, and the   prices offered by the other agents on the PCW. 
The agent does not know the exact    distributions of $(X_t,Y_t)_{t=1}^T$.
Instead, the agent has   access to 
an offline   dataset   consisting of      tuples $\mathcal{D}=(x_n, h_n, a_n, y_n)_{n=1}^N$
for $N$-customers, 
where the components 
correspond to the customer features,  quantile(s) of market prices quoted by other agents, historically quoted prices/actions taken, and customer decision, respectively.
In our particular instance, only around $2\%$ of $(y_n)_{n=1}^N$ are non-zero. This results in a low signal-to-noise ratio, and creates a sparse reward in the objective function \eqref{eq: RL objective}.

\begin{Remark}
    In \eqref{eq: RL objective}, we assume that the agent chooses the policy   without considering  constraints on the insurance portfolio. In practice,   the pricing policy may also depend on the 
    number of contracts already issued  to a     customer segment, target conversion rates and average premiums accepted.
Let's take the example of an insurer wanting to cap the total number of accepted policies  
 over the period $T$.
In this case,  the cumulative reward can be modified  into 
    \begin{equation*}
    \sE \left[ \sum_{t=1}^T Y_t\, r(X_t,\phi(t, X_t) ) + g(S_T)
    \right],
\end{equation*}
where $S_T\in \sR^{\cX}$ represents the 
the   number of converted quotes  over $T$ customers,
and  $g: \sR^{\cX}\to \sR$  represents the agent's preference  for the target portfolio.
The methodology developed herein can be easily adapted to this setting.
\end{Remark}

\section{Proposed Hybrid Reinforcement Learning methodology}
\label{sec:RL_methodology}

This section presents a RL framework for solving the pricing problem using the historical dataset $\mathcal{D}$. Since one cannot directly interact with the true environment, there are several challenges that need to be addressed. These include 1) creating an interactive environment for training   RL algorithms, 
2) designing a pricing system that can handle the non-stationarity of the market, and 
3) assessing the performance of a RL agent offline.
We mitigate these challenges  by integrating model-based and model-free methods into a \textbf{hybrid methodology}.

\paragraph{Contextual bandit problem for insurance pricing.} 
We begin by noting the objective in Eq.~\eqref{eq: RL objective} is the classic RL objective with discount factor~\cite{sutton2018reinforcement} set to $1$.
To facilitate  training  an agent, 
we  make the following assumptions on  the random variables 
$(X_t,Y_t)_{t=1}^T$ in \eqref{eq: RL objective}:
\begin{enumerate}[(i)]
    \item The customer   features $(X_t)_{t=1}^T$ are independently  and identically distributed.
    \item 
For each $t\in \{1,\ldots, T\}$,        the market  quantile price(s)  $H_t$ 
 of all agents 
 takes values in a space $\cH$, 
 and follows the (conditional) distribution $\psi^h(\cdot|X_t)$. Note, multiple quantile prices, or average values, are often made available and used by pricing rules engines. This does not impact our methodology.
    \item 
    For each $t\in \{1,\ldots,T\}$,
    the customer's decision $Y_t$ is 
    a conditional Bernoulli random variable. 
    More precisely, for each $t\in \{1,\ldots, T\}$, 
let  
  $X_t\in \cX $  be the customer features, 
      $A_t = \phi(t,X_t)\in \cA $  be  the agent price action
      and  
      $H_t\in \cH$  be       the market  quantile price(s) 
 of other agents (this is unknown in live but can be used for offline training). 
Then   $Y_t=1$  
  with probability $p(X_t,H_t,A_t)$ and $0$ otherwise,
where    $p:\cX\times \cA\times \cH\to [0,1]$
is 
 a deterministic function independent of $t$. 
  The function $p$ is often referred to as the customer
 \textbf{conversion model}. 

\end{enumerate}
 
These modelling assumptions imply that 
 it suffices to  find   a time-independent policy $\phi^\star :\cX\to \cA$ that maximises  the one-step reward:
\begin{equation}
\label{eq: RL objective_onestep}
\phi^\star\in \argmin_{\phi:\cX\to \cA}  \sE \left[ Y_t \cdot r(X_t, \phi(X_t) ) \right].
\end{equation}
This simplifies  
the optimisation problem \eqref{eq: RL objective_optimal}
into a CB problem.    The training dataset $\mathcal{D} $ consists of customers features $x$, market quantile price(s) $h$, the pricing action taken by the insurer $a$, along with the customer's decision $y$. 

To deploy an RL learning algorithm that solves \eqref{eq: RL objective_onestep} we require accessing  
customers' actions to  
prices quoted via the proposed training algorithm. Because training in the online setting (i.e using a deployed algorithm) would be prohibitively expensive (and risky), one needs to augment the training data set using simulations.   
To 
simulate the customer's response to a given quoted price,
we 
fix a   dataset 
 $\mathcal{D}_{\rm{train}}=(x_n, h_n, a_n, y_n)_{n=1}^{N_{\rm{train}}}\subset\mathcal{D}$,
 and 
estimate     a conversion function $p$  by  minimising a suitable loss function
$$
\hat{p} = \argmin_{p_\theta}\frac{1}{N_{\rm{train}}} \sum_{n=1}^{N_{\rm{train}}} \ell(y_n, p_{\theta}(x_n, h_n, a_n ) ),$$
over certain parametric models $p_{\theta} : \cX \times \cH \times \cA \to [0,1]$
such that    $a \mapsto p_{\theta}(x, h, a)$ is non-increasing for all $(x,h) \in \cX \times \cH$.  
This monotonicity constraint 
is applied to the estimated model $p_\theta$ as the true conversion probability $p(x,h,a)$ decreases as the quoted price $a$ increases. 

\begin{Remark}
Note that  although the market quantile price(s) is(are) dependent solely on the customer features, our conversion model incorporates both the customer feature and the market quantile price(s) explicitly. This is because customer behaviour, given quoted prices, is expected to be stable over a long period of time. By including the market quantile price(s) in our conversion model, we can fit the model using historical data over a relatively long time horizon and encode market dynamics into the agent via the simulations.

\end{Remark}

Given the fitted conversion  model $p_\theta$ using the data  $\mathcal{D}_{\rm{train}}=(x_n, h_n, a_n, y_n)_{n=1}^{N_{\rm{train}}}\subset\mathcal{D}$ (which may span a large time-horizon), we create a data set  $\tilde{\mathcal{D}}_{\rm{train}}$ based on more recent data points $(x_n,h_n)_{n\geq 1}$ to reflect current market conditions/customers. The construction is as follows:
\begin{enumerate}
    \item 
    At each time $t$,
  sample $(x, h)$ randomly from the  data set $ \tilde{\mathcal{D}}_{\rm{train}}$. 
    \item Submit a price from the pricing action $a$ generated from the agent's pricing rule $\phi: \cX\to \cA$.

    \item  Sample  $U \sim \text{Unif}(0,1)$, 
    and the agent observes  the customer decision defined by 
    $y = \bm{1}\big(U \leq \hat{p}(x,h,a) \big)$.
    
    \item The agent collects the reward, e.g., $y \,  r(x,a)$.
    
\end{enumerate}
This constitutes the simulator set-up used in training and evaluating agents, more details on our specific simulator environment(s) are provided in Section \ref{sec:experiments}.  

 \paragraph{Training RL algorithms with  dense reward.}

Due to the low conversion rate, the majority of simulated customer decisions will be zero, which makes it inefficient to train a RL/CB agent based solely on that binary outcome.
It also makes interpretability more difficult as variance in the simulated outcome may be a dominating factor during training.
Here we reformulate the problem
\eqref{eq: RL objective_onestep} to one
with dense rewards using the estimated conversion probability $\hat{p}$. 
We start by observing that for any given pricing rule $\phi:\cX\to \cA$,  by the tower's property of the conditional expectation, 
the reward in   \eqref{eq: RL objective_onestep} is equivalent to 
\begin{align}
\label{eq: RL modified objective}
\begin{split}
 \sE \left[ Y_t \cdot r(X_t, \phi(X_t) ) \right]
& =
\sE \left[ \sE[Y_t| X_t, H_t, \phi(X_t)] r(X_t, \phi(X_t) ) \right] 
\\
&  
= \sE \left[ p(X_t, H_t, \phi(X_t)) r(X_t, \phi(X_t) ) \right],
\end{split}
\end{align}
where $p:\cX\t \cH\t \cA\to [0,1]$ is the customer's true conversion probability, depending on the customer's features $X_t$, 
the market quantile price(s) $H_t$,
and the agent's quoted price/price action $\phi (X_t)$.
Substituting the true conversion model $p$ with the estimated model $\hat{p}$
yields the following approximation of    
\eqref{eq: RL modified objective}:
\begin{equation}
\label{eq: RL modified objective_estimate}
 \sE \left[ Y_t \cdot r(X_t, \phi(X_t) ) \right]
 \approx
 \sE \left[ \hat{p}(X_t, H_t, \phi(X_t)) r(X_t, \phi(X_t) ) \right].
\end{equation}
This  modified reward 
$\hat{R}_t  
= \hat{p}(X_t, H_t, \phi(X_t)) r(X_t, \phi(X_t) )
$ is non-zero for the majority of customer contexts/features and resolves the sparsity issue of the original reward design.

Based on the reformulation   \eqref{eq: RL modified objective_estimate}, 
many existing RL algorithms  can be employed to search the optimal policy.  
In the following, we 
present the actor-critic algorithm as an example, see Alg.~\ref{Alg: Online Actor-Critic}. The probabilistic actor policy $\pi_{\theta^a_m}$ from this algorithm is used as our pricing policy $\phi$  in live either directly or in a greedy-fashion by taking the argmax over the action space. 
Possible alternatives include 
the 
asynchronous 
advantage actor-critic (A3C) algorithm
\cite{li2017deep}, the TD3 algorithm \cite{fujimoto2018}, the proximal policy optimisation (PPO) algorithm \cite{schulman2017proximal} as well as simpler algorithms such as DQN~\cite{volodymyr2015} and its variants. 
 
\begin{algorithm}
\DontPrintSemicolon
\SetAlgoLined

  \KwInput{Iteration number $M$, a conversion model $\hat{p}$, a market  simulator, learning rates $(\gamma^{q}_m)_{m \in \sN}$ and $(\gamma^{a}_m)_{m \in \sN}$,
  }
 {
Choose  architectures  
$Q_{\theta^q}:\cX\times \cA\to \sR$
and 
$\pi_{\theta^a}:\cX\to \cP(\cA)$,
and 
initialise $\theta^a_1$ and $\theta^q_1$.
 }\;

  \For{$m = 1, 2,\ldots, M$}
 {
 {Generate a customer $x$ and the market quantile price $h$  via the simulator.}\;
{Sample $A \sim  \pi_{\theta^a_m}(\cdot | x)$.}\;
 {Agent collects a reward
 \begin{equation}
     \label{eq: RL reward actor-critic}
     R := \hat{p}(x,h,A) r(x,A).
 \end{equation}}\;
  {Update the critic  
$ 
     {\theta}^q_{m+1} = {\theta}^q_m - 2\gamma^{q}_m \Big(Q_{{{\theta}^q_m}}(x,a) - R \Big) \nabla_{\theta^q} Q_{{{\theta}^q_m}}(x,a).
$ }\;
 {Update the actor  
$
     {\theta}^a_{m+1} = {\theta}^a_m  + \gamma^a_m Q_{{{\theta}^q_m}}(x,a) \nabla_{{\theta}^a} \log \pi_{\theta^a_m}(A | x).
$
 }\;
}
 {\textbf{Return:}
   $\pi_{\theta^a_m} : \cX \to \cP(\cA)$.
  }\;

 \caption{Actor-Critic algorithm}
 \label{Alg: Online Actor-Critic}

\end{algorithm}

\paragraph{Summarising the hybrid methodology.} The approach described in this Section combines the benefits of both model-based and model-free methods. It utilises a model-based approach to transform the sparse reward into a dense reward, thereby improving the sample efficiency of fully model-free algorithms. At the same time, it employs a model-free approach to update the policy sequentially based on the customer features and simulated dense reward.
This approach differs from fully model-based algorithms where modelling the dependence of the market quantile price(s) on the customer features, i.e., $\psi^h$. This model-based approach has several drawbacks. Since the market quantile price(s) typically depend  nonlinearly   on the customer features, it is challenging to choose an appropriate architecture to accurately approximate it. Additionally, historical data indicates that modelling this relationship between customer features and aggregated prices can degrade in performance quickly, and are often rebuilt or recalibrated as regularly as monthly where possible. In contrast, the proposed hybrid method only requires quoted quantile price(s) $(h_t)_{t\ge 0}$, which can be dynamically updated as new data arrives with little additional complexity.
 %

\paragraph{Evaluating agent performance in a consistent manner.}

After a pricing policy has been trained, its performance  can be evaluated by constructing a new market simulator using a test dataset. However, when comparing pricing policies generated by different RL/CB algorithms using historical data, it is crucial to ensure consistency in customer behaviour. Specifically, it is important to ensure that if a customer accepts a quoted price, they will also accept any other quoted price that is lower. To achieve this, let $\phi_1$ and $\phi_2$ be two different pricing rules, and let $\hat{p}:\cX\times \cH\times \cA\to [0,1]$ be an estimated conversion model. Then, for a given customer feature $x$ and market quantile price $h$, the customer's decision under each pricing rule can be simulated by first sampling $U \sim \text{Unif}(0,1)$, and defining the decision as $y_i = \bm{1}\big(U \leq \hat{p}(x,h,\phi_i(x)) \big)$, $i=1,2$. Note that the two decisions are determined by the same random sample $U$, enabling a fair comparison among different algorithms being evaluated. In reality, customers may react randomly to quoted prices - this results in large variances and require lots of additional computational cost to determine if an algorithm is performing better than an alternative.

\section{Numerical experiments}
\label{sec:experiments}
To demonstrate our methdology, we generate synthetic dataset that is reflective of a private commercial PCW data made available to the authors\footnote{Data provided by Esure Group.}. Using this synthetic data, we compare a series of agents vs  an actor-critic trained offline using our hybrid methodology. We now discuss the construction of this data, along with the underlying assumptions about the real environment that informs it, before describing the set-up and results of our numerical experiments. 

\begin{figure}[h]
	\centering
	\includegraphics[width=0.5\linewidth]{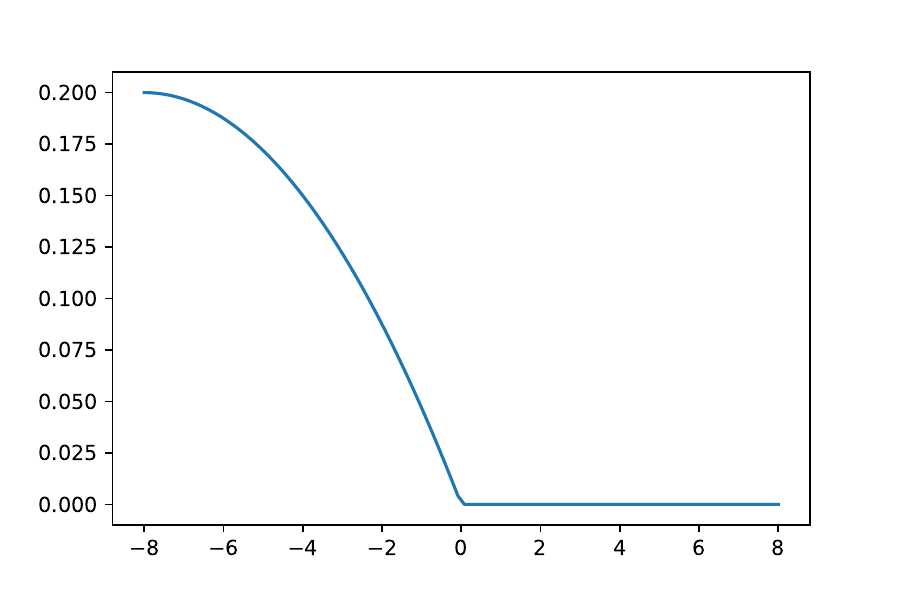}
	\hspace{-4mm}
	\includegraphics[width=0.5\linewidth]{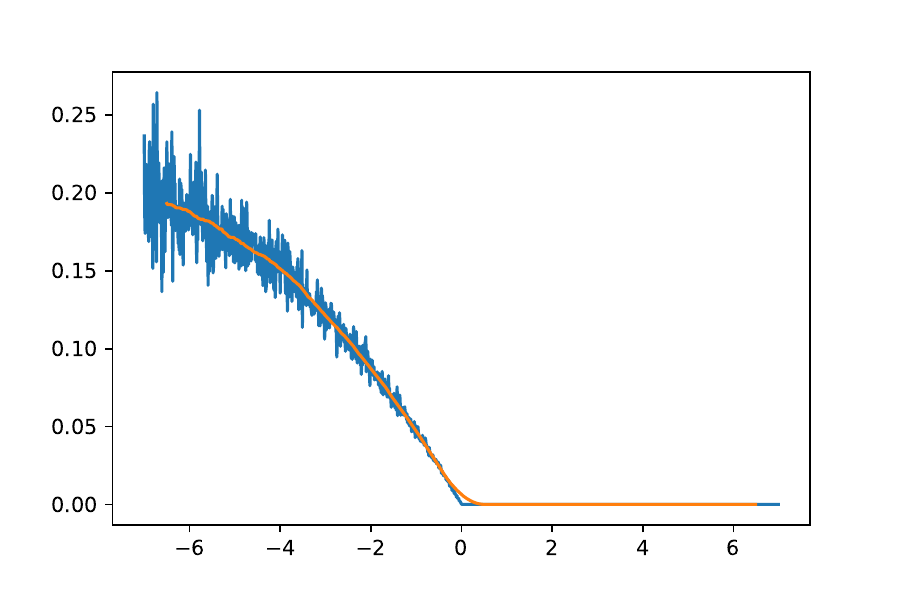}
	\caption{Exact and estimated conversion models.
	Left: The exact conversion probability as a function of the normalised price;
	Right: Comparison between the empirical  conversion model (blue curve) and the estimated
 conversion model (orange curve) for different normalised prices.} 
	\label{fig: True_demand}
\end{figure}

\paragraph{Understanding how to simulate the market.}
Analysing real-world PCW data, we identified the key relationship determining whether a customer converted was between the final quoted premium ($P$), the average top 5 price they received and the average prices of the insurers ranked 6-10 on the PCW. The latter two quantities are the market quantile prices discussed throughout and we denote them by Avg. Top5 and Avg. Top6-10 respectively. The final quoted premium was normalised using this market quantile information to produce a normalised price $z$ as follows:
\begin{equation}
\label{eq:regressor}
\text{z} = \frac{P - \text{Avg. Top5}}{\text{Avg. Top6-10} - \text{Avg. Top5}}.
\end{equation} 
This normalised price measures how competitive our quoted premium is relative to our competitors in the market. From analysing the real-world data, the conversion probability given this normalised price, $p(z)$, was found to follow this distribution:
\begin{equation}
\label{eq:conversion_model}
p(z) =
\begin{cases}
    0.2 &: z < -8, \\
    -0.2(z/8+1)^2+0.2 &: -8 \leq z < 0, \\
    0 &: z \geq 0.
\end{cases}
\end{equation}
The plot of this demand, $p(z)$, is shown Figure \ref{fig: True_demand} (left).
We conclude this portion by noting the core assumption derived from analysing the PCW data is that our expected true conversion model depends primarily on the normalised price $z$, independent of other customer features. These features provide higher order corrections to the model but are not the primary drivers given the actual market quantile prices.

\paragraph{Construction of synthetic dataset.}
A synthetic dataset of $35000$ customers on a PCW, with $16$ customer features (spanning age to credit risk data) along with simulated Avg. Top5 prices for each of these customers was provided for direct modelling. The Avg. Top5 prices provided are derived from these $16$ customer features. The distributions of these features and market price were created to reflect realistic PCW data.

This dataset was augmented to generate a benchmark premium ($P_0$),  Avg. Top6-10 and "burn" cost ($b$). This augmentation was done by modifying the Avg. Top5 price provided and hence depends indirectly on the customer features, $x$. This augmentation was performed via the use of scaling factors randomly sampled from normal distributions. The scaling factors were chosen to reflect the magnitude of the conversion rate of the original dataset and the shape of the demand curve.

Comparing to real data, we found the following calculation reflected these real-world constraints:
\begin{align*}
    b(x) &= \text{Avg. Top5} \times N(0.8, 0.2), \\
    P_0(x) &= \text{Avg. Top5} \times N(1, 0.1), \\
    \text{Avg. Top6-10} &= \text{Avg. Top5} \times |N(1, 0.3)|,
\end{align*}
where $N(\mu,\sigma)$ denotes a normal random variable with mean $\mu$ and standard deviation $\sigma$.

The synthetic final quoted premium $P$ will then be generated as follows:
\begin{equation}
P = \text{Avg. Top5} \times N(1, 0.3).
\end{equation}

From these parameters, we can define the observed action taken by the insurer to be the ratio $\frac{P}{P_0}$. We abuse our notation slightly to denote this ratio as a price action $a$ with the reward function $r$, for a customer with features $x$, \eqref{eq: RL objective_optimal} then being given by:
\begin{equation}
r(x,a) = a\times P_0(x) - b(x) = P(x,a) - b(x).
\end{equation}
This is effectively the profit in year one, i.e., the CLTV1 for this customer.

We split the initial $35000$ customers into two sets - $28000$ in train and $7000$ in test. We then generate  $5 \times 10^6$ samples (with replacement) from the training data set.  For each sample, the premium $P$ is generated  and the customer decision is determined via the conversion probability $p(z)$.
This leads to  a complete training data set $\mathcal{D}_{\textrm{train}}$ with $5 \times 10^6$ data points. Using this dataset we construct a training simulator. 

\paragraph{Construction of the training environment.}
We now construct  the training environment based on the training data set, without using $p(z)$ nor the testing data set.
This represents how one can use their own conversion data to create a market simulator directly from the data.

We start by {estimating} the demand curve from the data.
We split the training data  $\mathcal{D}_{\textrm{train}}$ (of size $5 \times 10^6$) into bins according to the size of the 
$z$ rounded to the second decimal place.

Let $p_R$ be
the empirical  conversion probability within the $R$-th bin.
The estimated conversion model $\hat{p}$ is then defined by 
\begin{equation}
    \label{eq: conversion estimate}
\begin{aligned}
    \tilde{p}_R &= \frac{1}{100}\sum_{n=1}^{100} p_{R+0.01n -0.5}, \quad 
    \hat{p}_R &= \min(\hat{p}_{R-0.01}, \tilde{p}_R),
\end{aligned}
\end{equation}
where we first smooth the empirical  conversion probability
using a moving average, and then take the minimum between consecutive bins to ensure the estimated model decreases with respect to the normalised price.
For simplicity, we extrapolate the estimated model outside the support of training data, by setting  $\hat{p}_R = \hat{p}_{-6}$ for $R < -6$ and  $\hat{p}_R = \hat{p}_{6}$ for $R > 6$.
The function  $\hat{p}_R$ will be used  as an estimated conversion probability to generate customer decisions in the training environment.   Figure  \ref{fig: True_demand}(right) 
shows $p_R$ (in blue) and $\hat{p}_R$ (in orange) for $R \in [-6,6]$. In practice, for a given $z$ this is mapped to a bucket $R$ and $\hat{p}_R$ represents the expected conversion for that normalised price. This will be denoted $\hat{p}(z)$ from the coming paragraphs to emphasise that it is a mapping from $P$ (and hence $z$) to an estimated conversion with the bins serving as a useful intermediate grouping for the model.


\paragraph{Training RL/CB Agents.} Using the estimated conversion, $\hat{p}(z)$, we train two types of agent. Both agents are trained in the same manner, using the actor-critic algorithm presented in \ref{Alg: Online Actor-Critic}, but they differ in their reward function. The first agent is a standard RL/CB agent trained using a sparse reward while the second uses our {hybrid} methodology with the associated dense reward. At each iteration of the training process a customer - along with their features $x$, Avg. Top5, Avg. Top6-10, $P_0$ and $b$ are sampled. The agents take price scaling actions. These actions are in the range $[0.7 - 1.3]$ and are discretised into $600$ actions where the separation is $0.001$ between each action, i.e.,  $a \in \{0.7, 0.7001, ..., 1.2999, 1.3\}$. This range of actions reflects the range of price scalings an insurer would typically consider, to scale up $P_0$ to the final quoted premium $P$. 

For the \textbf{standard RL agent}, the reward function for this customer at iteration $t\ge1$, is given by:
\begin{equation}
    \label{eq: reward sparse}
r(x_t,a_t) = Y_t \times (a_t \times P_0(x_t)-b(x_t)),
\end{equation}
where $Y_t$ is the customer decision simulated from $\hat{p}(z_t)$ and $x_t$, $a_t$ and $z_t$ are the customer features, agent action and the normalised price at time $t$.

The second \textbf{hybrid RL agent} incorporates the estimated conversion model
in the reward. In this case, it updates the policy using the following reward:
 \begin{equation}
    \label{eq: reward dense}
    r(x_t, a_t) =  \hat{p}(z_t) \times (a_t \times P_0(x_t)- b(x_t)).
\end{equation}

\paragraph{Benchmark pricing strategies.} 
 In addition to these agents, we also consider several benchmark pricing strategies to compare the RL/CB agents against. 
 The actor-critic agents cannot obtain the Avg. Top5 and the Avg. Top6-10 values from the data when providing prices and instead indirectly infer them through the reward. As a benchmark, we compare them to model-based agents who've access to these quantities with some small systematic errors. Specifically, we consider the following three scenarios: 

\textbf{Unbiased Estimation -}   the agent estimates the market by: 
\begin{align*}
    \text{Unbiased estimated average top 5} &= \text{Avg. Top5} \times N(1, 0.3) \\
    \text{Unbiased estimated average top 6-10} &= \text{Avg. Top6-10} \times N(1, 0.3)
\end{align*}

\textbf{Over Estimation -}  the agent estimates the market by: 
\begin{align*}
    \text{Over-estimated average top 5} &= \text{Avg. Top5} \times N(1.2, 0.3) \\
    \text{Over-estimated  average top 6-10} &= \text{Avg. Top6-10} \times N(1.2, 0.3)
\end{align*}

\textbf{Under Estimation -}  the agent estimates the market by: 
\begin{align*}
    \text{Under-estimated average top 5} &= \text{Avg. Top5} \times N(0.8, 0.3) \\
    \text{Under-estimated average top 6-10} &= \text{Avg. Top6-10} \times N(0.8, 0.3)
\end{align*}

These scenarios define an additional $3$ model-based agents to compare against. In all instances, the agents compute an estimate of the normalised price $\tilde{z}$ in accordance with Eq.~\eqref{eq:regressor} but using their estimated market Avg. Top5 and Avg. Top6-10 values in place of the actual values, that is:
$$
    \tilde{z} = \frac{P - \text{Estimated Avg. Top5}}{\text{Estimated Avg. Top6-10} - \text{Estimated Avg. Top5}}.
$$
Using this estimated normalised price and estimated conversion $\hat{p}$, the final price offered at time $t$ is yielded via maximising the map:
$$P\mapsto \hat{p}(\tilde{z})\times (P - b(x_t)).$$

Finally, we consider $2$ extreme pricing rules. 
We consider a random agent that quotes a price generated by randomly selecting from the action set. This random agent serves as a worst-case benchmark to evaluate the performance of other pricing policies. We will also consider a perfect information agent where
 at each time $t$, this agent  maximises the expected reward:
$$P\mapsto p(z)  \times (P - b(x_t)),$$
using  the true conversion model $p$  defined in \eqref{eq:conversion_model},
and the actual Avg. Top5 and Avg. Top6-10 market prices.
Although this perfect information agent cannot be implemented in practice  (since  the exact customer conversion model is unknown as are the market quantile prices in live),
it serves as the best-case benchmark for the proposed hybrid RL agent.  In conclusion, we have $7$ agents to evaluated - $6$ along with our hybrid agent.

\paragraph{Performance evaluation of  RL agents and results.}

Let $\phi_i$, $i=1,2,\ldots, 7$, denote the pricing rules generated by the standard RL agent, the hybrid RL agent, the unbiased model-based agent, the over-estimated model-based agent, the under-estimated model-based agent, the random agent and the perfect-information agent, respectively.

The following Algorithm  \ref{Alg: PerfEval} summarises the procedure to evaluate the performance of these pricing rules using a test data set $\mathcal{D}_{\rm test}$ and the true conversion model $p(z)$ in \eqref{eq:conversion_model}.

\begin{algorithm}
\DontPrintSemicolon
\SetAlgoLined
 \KwInput{Pricing rules $(\phi_i)_{i=1}^7$, true conversion model $p$ for the market, testing data set $\mathcal{D}_{\rm test}$ of size $N_{\rm test}$.}
 {

	\For{$t = 1, 2,\ldots, N_{\rm test}$}
 	{
 		{Sample an entry from $\mathcal{D}_{\rm test}$, containing the customer features ($x_t$), the Avg. Top5 and Avg. Top6-10 prices, and the cost $b$ given the customer features.}\;
		{All agents quote their prices, denoted by $(P_i)_{i=1}^7$. The perfect information agent uses the entire entry  and the exact conversion model, and the other agents only uses 	the customer feature.}\;
		{Sample $U \sim U[0, 1]$ and store this value.}\;

	\For{$i = 1,\ldots,7$}
		{

 			{Compute the normalised price $z_i$ from $P_i$ along with the associated conversion probability $p_i = p(z_i)$ where $p$ is defined in  \eqref{eq:conversion_model}.} \;
 			{Record the expected reward, $p_i \times (P_i - b(x_t))$.}\;
		{Record the realised reward, $\bm{1}(U <  p_i) \times (P_i - b(x_t))$.}\;
		}
	}
}
 \caption{Performance evaluation}
 \label{Alg: PerfEval}
\end{algorithm}


\begin{figure}[ht!]
	\centering
	\includegraphics[width=0.5\linewidth]{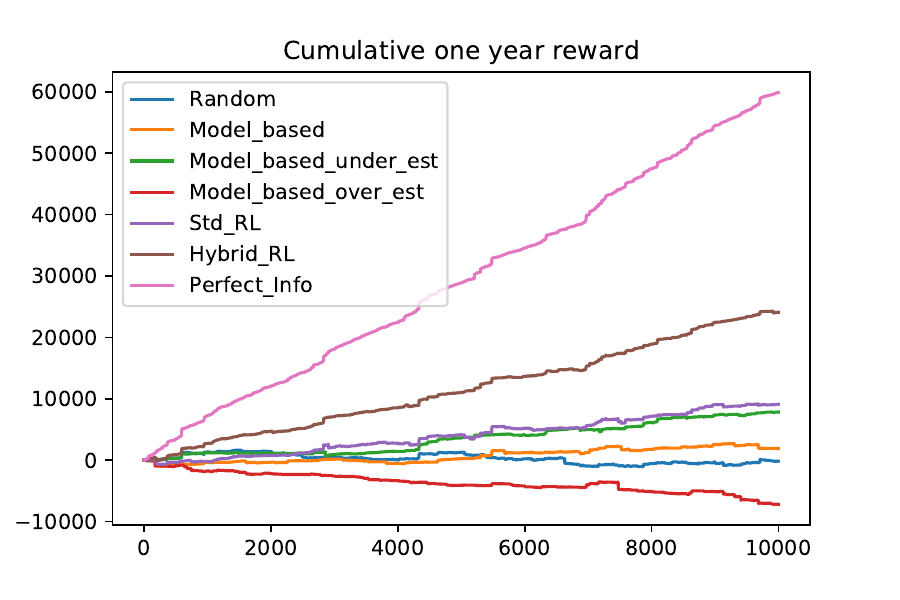}
	\hspace{-4mm}
	\includegraphics[width=0.5\linewidth]{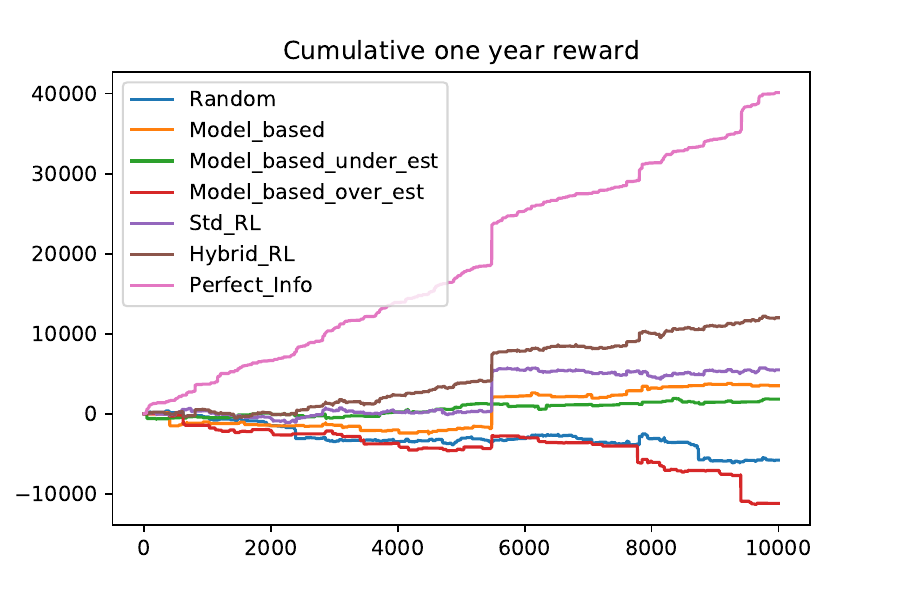}
	\caption{Comparison of the expected (left) and realised (right) cumulative  rewards among all agents.}
	\label{fig:exp reward}
\end{figure}


Figure \ref{fig:exp reward}
compare the cumulative  expected rewards and the cumulative realised reward 
for all agents.  Although the proposed hybrid RL (brown line) underperforms the (unrealistic) perfect information agent,
it clearly outperforms the remaining $5$ agents, including the standard model-free RL agent (purple line). Unsurprisingly, the random agent makes losses very quickly. This highlights the need to avoid highly exploratory behaviour that frequently occurs at the early-stages of training an online RL agent. Furthermore, we observe the rate at which the hybrid agent accumulates reward outperforms all other agents, again with the exception of the perfect information agent. This demonstrates the improved sample efficiency from use of our hybrid approach compared to more traditional pricing strategies/benchmarks.

\section{Conclusions and Future Work}
\label{sec:conclusions}
In this paper, we formulated the problem of pricing insurance at new business on a price comparison site as a RL   problem. We addressed the cold-start problem, interpretability, reward sparseness and partial observability of a non-stationary market by 
integrating 
model-based and model-free RL methods. The model-based component creates a dense interpretable reward  and allows for the creation of an effective market simulator. This simulator allows us to train model-free RL/CB methods offline, with these methods learning the market dynamics implicitly from the simulator by-passing both the cold-start and observability issues when run in live.

We evaluated this approach on representative synthetic data derived from real-world PCW data.  Both the expected and realised CLTV1 for the  {hybrid} agent performed better than the benchmarks, with the exception of the (unrealistic)  agent which has perfect information on the market environment. Our approach is readily extendible to scenarios where the constraints on the agent's behaviour, such as the rate of conversion, apply. This can be achieved via an adjustment in the reward function. We expect future work in this area to examine (i) the potential use of 'Multi-objective RL' techniques where an agent has multiple competing objectives~\cite{hayes2022,yang2019} and constraints. (ii) Incorporate uncertainty estimates into the agent and its exploration policy~\cite{ghavamzadeh2015,green2021}. (iii) Improvements to the market simulator with direct incorporation of time-dynamics, usage of agent based simulation~\cite{lin2011,fraunholz2021} as well as an associated empirical study on the effectiveness of various RL/CB algorithms with this simulator.

\section{Authors' Contributions}
 All authors conceptualized the study. 
 TT, YZ and LS proposed the methodology,
 and TT conducted the numerical experiments. 
 TT, YZ and LS  wrote the manuscript,
and   all authors critically revised the manuscript.   

\bibliographystyle{abbrv}
\bibliography{insurance_RL.bib}

\begin{thebibliography}{10}

\bibitem{ban2021personalized}
G.-Y. Ban and N.~B. Keskin.
\newblock Personalized dynamic pricing with machine learning: High-dimensional
  features and heterogeneous elasticity.
\newblock {\em Management Science}, 67(9):5549--5568, 2021.

\bibitem{blier2020machine}
C.~Blier-Wong, H.~Cossette, L.~Lamontagne, and E.~Marceau.
\newblock Machine learning in {P\&C} insurance: A review for pricing and
  reserving.
\newblock {\em Risks}, 9(1):4, 2020.

\bibitem{cai2017}
H.~Cai, K.~Ren, W.~Zhang, K.~Malialis, J.~Wang, Y.~Yu, and D.~Guo.
\newblock Real-time bidding by reinforcement learning in display advertising.
\newblock In {\em Proceedings of the Tenth ACM International Conference on Web
  Search and Data Mining}, WSDM '17, page 661–670, New York, NY, USA, 2017.
  Association for Computing Machinery.

\bibitem{collier2018}
M.~Collier and H.~U. Llorens.
\newblock Deep contextual multi-armed bandits.
\newblock {\em CoRR}, abs/1807.09809, 2018.

\bibitem{devidze2021}
R.~Devidze, G.~Radanovic, P.~Kamalaruban, and A.~K. Singla.
\newblock Explicable reward design for reinforcement learning agents.
\newblock In {\em Neural Information Processing Systems}, 2021.

\bibitem{eschmann2021}
J.~Eschmann.
\newblock {\em Reward Function Design in Reinforcement Learning}, pages 25--33.
\newblock Springer International Publishing, Cham, 2021.

\bibitem{fraunholz2021}
C.~Fraunholz, E.~Kraft, D.~Keles, and W.~Fichtner.
\newblock Advanced price forecasting in agent-based electricity market
  simulation.
\newblock {\em Applied Energy}, 290:116688, 2021.

\bibitem{fujimoto2019}
S.~Fujimoto, D.~Meger, and D.~Precup.
\newblock Off-policy deep reinforcement learning without exploration.
\newblock In K.~Chaudhuri and R.~Salakhutdinov, editors, {\em Proceedings of
  the 36th International Conference on Machine Learning}, volume~97 of {\em
  Proceedings of Machine Learning Research}, pages 2052--2062. PMLR, 09--15 Jun
  2019.

\bibitem{fujimoto2018}
S.~Fujimoto, H.~van Hoof, and D.~Meger.
\newblock Addressing function approximation error in actor-critic methods.
\newblock In J.~G. Dy and A.~Krause, editors, {\em Proceedings of the 35th
  International Conference on Machine Learning, {ICML} 2018,
  Stockholmsm{\"{a}}ssan, Stockholm, Sweden, July 10-15, 2018}, volume~80 of
  {\em Proceedings of Machine Learning Research}, pages 1582--1591. {PMLR},
  2018.

\bibitem{ghavamzadeh2015}
M.~Ghavamzadeh, S.~Mannor, J.~Pineau, and A.~Tamar.
\newblock Bayesian reinforcement learning: A survey.
\newblock {\em Found. Trends Mach. Learn.}, 8(5–6):359–483, nov 2015.

\bibitem{green2021}
R.~Green, M.~Rowe, and A.~Polleri.
\newblock Macest: The reliable and trustworthy model agnostic confidence
  estimator.
\newblock {\em CoRR}, abs/2109.01531, 2021.

\bibitem{guelman2014}
L.~Guelman and M.~Guillen.
\newblock A causal inference approach to measure price elasticity in automobile
  insurance.
\newblock {\em Expert Systems with Applications}, 41(2):387--396, 2014.

\bibitem{han2021}
B.~Han and C.~Arndt.
\newblock Budget allocation as a multi-agent system of contextual \& continuous
  bandits.
\newblock In {\em Proceedings of the 27th ACM SIGKDD Conference on Knowledge
  Discovery \& Data Mining}, KDD '21, page 2937–2945, New York, NY, USA,
  2021. Association for Computing Machinery.

\bibitem{hayes2022}
C.~F. Hayes, R.~R\u{a}dulescu, E.~Bargiacchi, J.~K\"{a}llstr\"{o}m,
  M.~Macfarlane, M.~Reymond, T.~Verstraeten, L.~M. Zintgraf, R.~Dazeley,
  F.~Heintz, E.~Howley, A.~A. Irissappane, P.~Mannion, A.~Now\'{e}, G.~Ramos,
  M.~Restelli, P.~Vamplew, and D.~M. Roijers.
\newblock A practical guide to multi-objective reinforcement learning and
  planning.
\newblock {\em Autonomous Agents and Multi-Agent Systems}, 36(1), apr 2022.

\bibitem{khraishi2022offline}
R.~Khraishi and R.~Okhrati.
\newblock Offline deep reinforcement learning for dynamic pricing of consumer
  credit.
\newblock In {\em Proceedings of the Third ACM International Conference on AI
  in Finance}, pages 325--333, 2022.

\bibitem{krashen2019}
E.~Krasheninnikova, J.~Garcia, R.~Maestre, and F.~Fernandez.
\newblock Reinforcement learning for pricing strategy optimization in the
  insurance industry.
\newblock {\em Engineering Applications of Artificial Intelligence}, 80:8--19,
  2019.

\bibitem{li2017deep}
Y.~Li.
\newblock Deep reinforcement learning: An overview.
\newblock {\em arXiv preprint arXiv:1701.07274}, 2017.

\bibitem{liu2022}
M.~Liu, J.~Liu, Z.~Hu, Y.~Ge, and X.~Nie.
\newblock Bid optimization using maximum entropy reinforcement learning.
\newblock {\em Neurocomputing}, 501:529--543, 2022.

\bibitem{lundberg2017}
S.~M. Lundberg and S.-I. Lee.
\newblock A unified approach to interpreting model predictions.
\newblock In {\em Proceedings of the 31st International Conference on Neural
  Information Processing Systems}, NIPS'17, page 4768–4777, Red Hook, NY,
  USA, 2017. Curran Associates Inc.

\bibitem{luo2023distribution}
Y.~Luo, W.~W. Sun, and Y.~Liu.
\newblock Distribution-free contextual dynamic pricing.
\newblock {\em Mathematics of Operations Research}, 2023.

\bibitem{volodymyr2015}
V.~Mnih, K.~Kavukcuoglu, D.~Silver, A.~A. Rusu, J.~Veness, M.~G. Bellemare,
  A.~Graves, M.~A. Riedmiller, A.~Fidjeland, G.~Ostrovski, S.~Petersen,
  C.~Beattie, A.~Sadik, I.~Antonoglou, H.~King, D.~Kumaran, D.~Wierstra,
  S.~Legg, and D.~Hassabis.
\newblock Human-level control through deep reinforcement learning.
\newblock {\em Nat.}, 518(7540):529--533, 2015.

\bibitem{prudencio2023}
R.~F. Prudencio, M.~R. O.~A. Maximo, and E.~L. Colombini.
\newblock A survey on offline reinforcement learning: Taxonomy, review, and
  open problems.
\newblock {\em IEEE Transactions on Neural Networks and Learning Systems},
  pages 1--0, 2023.

\bibitem{stable-baselines3}
A.~Raffin, A.~Hill, A.~Gleave, A.~Kanervisto, M.~Ernestus, and N.~Dormann.
\newblock Stable-baselines3: Reliable reinforcement learning implementations.
\newblock {\em Journal of Machine Learning Research}, 22(268):1--8, 2021.

\bibitem{raju2006learning}
C.~Raju, Y.~Narahari, and K.~Ravikumar.
\newblock Learning dynamic prices in electronic retail markets with customer
  segmentation.
\newblock {\em Annals of Operations Research}, 143:59--75, 2006.

\bibitem{rana2015dynamic}
R.~Rana and F.~S. Oliveira.
\newblock Dynamic pricing policies for interdependent perishable products or
  services using reinforcement learning.
\newblock {\em Expert Systems with Applications}, 42(1):426--436, 2015.

\bibitem{saleem2022}
R.~Saleem, B.~Yuan, F.~Kurugollu, A.~Anjum, and L.~Liu.
\newblock Explaining deep neural networks: A survey on the global
  interpretation methods.
\newblock {\em Neurocomputing}, 513:165--180, 2022.

\bibitem{schulman2017proximal}
J.~Schulman, F.~Wolski, P.~Dhariwal, A.~Radford, and O.~Klimov.
\newblock Proximal policy optimization algorithms.
\newblock {\em arXiv preprint arXiv:1707.06347}, 2017.

\bibitem{shrikumar2017}
A.~Shrikumar, P.~Greenside, and A.~Kundaje.
\newblock Learning important features through propagating activation
  differences.
\newblock {\em CoRR}, abs/1704.02685, 2017.

\bibitem{sutton2018reinforcement}
R.~S. Sutton and A.~G. Barto.
\newblock {\em Reinforcement learning: An introduction}.
\newblock MIT press, 2018.

\bibitem{verschuren2022}
R.~M. Verschuren.
\newblock Customer price sensitivities in competitive insurance markets.
\newblock {\em Expert Systems with Applications}, 202:117133, 2022.

\bibitem{yang2019}
R.~Yang, X.~Sun, and K.~Narasimhan.
\newblock {\em A Generalized Algorithm for Multi-Objective Reinforcement
  Learning and Policy Adaptation}.
\newblock Curran Associates Inc., Red Hook, NY, USA, 2019.

\bibitem{lin2011}
C.~yu~Lin, N.~H. Kilicay-Ergin, and G.~E. Okudan.
\newblock Agent-based modeling of dynamic pricing scenarios to optimize
  multiple-generation product lines with cannibalization.
\newblock {\em Procedia Computer Science}, 6:311--316, 2011.
\newblock Complex adaptive sysytems.

\end{thebibliography}
\end{document}